\documentclass[twocolumn,twoside,slac]{revtex4}
\usepackage{graphicx}
\usepackage{fancyhdr}
\begin{document}

\title{Data Acquisition Software for CMS HCAL Testbeams}


\author{J. Mans and W. Fisher}
\affiliation{Princeton University, Princeton, NJ 08544, USA}

\begin{abstract}
Although CMS will not start operation for several years, many subdetector
groups have active testbeam programs involving final or near-final electronics.
The high-bandwidth electronics require the development of
new DAQ systems including hardware and software.  This paper presents the
design and performance of the DAQ software used for the summer 2002 HCAL
testbeam at CERN.  This system successfully acquired over 80M testbeam events,
running at rate of 200-1000 Hz.  The paper also describes some of the developments
made for the 2003 testbeam, including the integration of
the CMS xDAQ framework.
\end{abstract}

\maketitle

\thispagestyle{fancy}

\section{Introduction}

The CMS experiment is a large general-purpose high energy physics detector
which will be used to study proton-proton and heavy-ion collisions at the 
Large Hadron Collider (LHC).  Although the experiment will not start operation
for several years, the groups developing portions of the experiment, or 
subdetectors, are actively testing the electronics which will be used in the 
final experiment.  The Hadron Calorimeter (HCAL) group is well along in this
project, and the current testbeam program is focused on evaluating the
final designs for the electronics.  

The HCAL testing program requires the
development of data acquisition software which can handle the large number of
channels and high data rate which the final electronics can produce.
This paper describes the data acquisition software which was written for the
2002 test beam carried out at CERN, as well as describing the changes and 
extensions made for the 2003 program.

\subsection{Review of the HCAL Data Chain}

\begin{figure}[ht]
\begin{center}
\includegraphics[width=\linewidth]{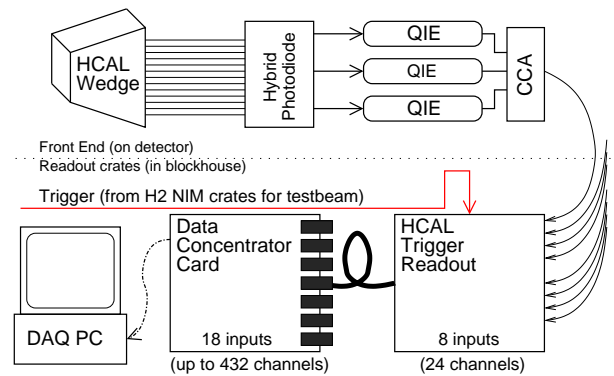}
\end{center}
\caption{ Schematic depiction of the HCAL Testbeam 2002 readout chain from HCAL wedge to 
DAQ PC.  The CCA and HTR are connected by a 1.4 Gbaud link optical link.  The HTR is connected
to the DCC via a 10-conductor copper LVDS link.  The DCC transmits events to the DAQ PC over
32-bit SLINK (optical fiber).
\label{fig-chain} }
\end{figure}

Before launching into the details of the testbeam DAQ, it is worthwhile to
briefly review CMS HCAL and its the data readout chain.  
Here we describe the setup used 
for the 2002 testbeam.  The readout chain is shown schematically in
Figure~\ref{fig-chain}.

The CMS HCAL is a sampling calorimeter consisting of a brass absorber with
plastic scintillator panels inserted into the absorber.  The scintillation light 
from the panels is collected by wavelength-shifting fibers and transported to
a hybrid photodiode (HPD) which converts the light into electrical charge.
The charge signal is measured and encoded into a non-linear digital
scale by the Charge Integrator IC (QIE).  The QIE uses the LHC clock to 
divide time into regular bins and measures the accumulated charge in each time
bin.  Internally, the QIE uses capacitors to accumulate the charge and measure
the voltage.  There are four such capacitors in each QIE, and the QIE uses
each capacitor in turn, discharging it for two clocks before using it again.
Thus each subsequent time sample comes from a different capacitor-id or CAPID.
In the final system, each time bin will be 25ns long, but for the 2002
testbeam a slower clock was used, which made each bin 33ns long.

The outputs of three QIE channels are digitally combined onto a high-speed
optical link and sent to the HCAL Trigger/Readout (HTR) board \cite{htr}.  
The 2002 HTR board could accept eight fiber links, corresponding to twenty-four QIE 
channels.  
The HTR board buffers the incoming digital data and transfers it to the Data 
Concentrator (DCC) \cite{dcc} when a Level 1 Accept (L1A) is received.  
Future revisions of the HTR board will also calculate trigger primitives for the 
trigger decision.
The DCC is responsible for collating data from up to eighteen HTRs 
and transferring it on to the central DAQ.  In the 2002 testbeam, the DCC transferred
the data over a 32-bit SLINK fiber link to a PC.  

\subsection{Requirements}

The testbeam DAQ has several important requirements, which are not necessarily
identical to the requirements for the final DAQ.  

\begin{figure}[ht]
\begin{center}
\includegraphics[width=\linewidth]{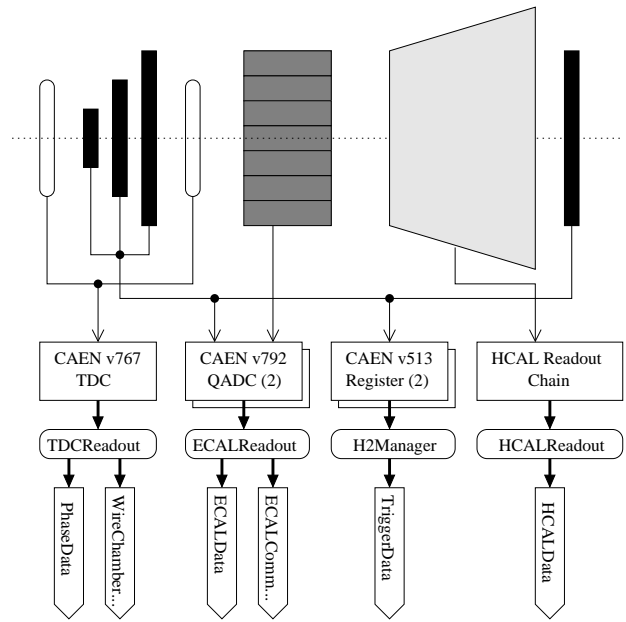}
\end{center}
\caption{ Schematic layout of the subdetectors used in the HCAL 2002 Testbeam.  The dark 
rectangles represent S1, S2, and S3 trigger scintillator counters and the muon counter.  The rounded
rectangles represent the wire chambers, while the dark gray rectangles represent the ECAL and the
wedge the HCAL.  Each subdetector is linked to the electronics used to read it out.  Each group
of electronics is coupled to the software module which acquires the data from the electronics.  The
tags at the bottom of the diagram indicate which final data streams are produced by which software
readout modules.
\label{fig-readout} }
\end{figure}

\begin{itemize}
\item {\bf Interfacing with testbeam trigger} -- the trigger electronics used
for the 2002 testbeam provide information to the DAQ through a CAENV592 
VME NIM register and require feedback from the DAQ through a second CAENV592.

\item {\bf Readout of HCAL data input chain} -- the HCAL data chain 
generates 976 bytes of data per HTR per event or a total of 5856 bytes for
each testbeam event (with six HTRs running).  This data is dumped into a DMA
buffer and the DAQ must parse and store it.
\item {\bf Readout of testbeam subdetectors} -- the testbeam setup for the HCAL
in 2002 includes several additional subdetectors such as a seven-by-seven 
crystal ECAL mock-up readout with CAENV792 QADCs and two wire chambers 
readout with a CAENV767 TDC.  A schematic showing the relationship between the
subdetectors and their readout electronics is shown in Figure~\ref{fig-readout}.
\item {\bf Flexible design} -- the testbeam DAQ should be flexible enough to
function without the testbeam trigger system for special runs or testing
away from the H2 testbeam area at CERN.  Only minor configuration changes 
should be required to enable data taking with different sets of subdetectors or
trigger sources.
\item {\bf Radioactive source calibration} -- the DAQ must also be able to
handle data from the radioactive source calibration, where the HTRs produce
histograms of data from the different QIEs and transmit them at a regular rate
without triggers.  The DAQ must sift through the DMA buffers and find them.
\end{itemize}

\section{Structure of the DAQ Software}

\subsection{Modules}

The DAQ software is written in C++ and runs under Linux kernels 2.2 and 2.4.  
The DAQ is divided into a set of modules, each of which is a separate thread 
of execution, running under the pthreads\cite{pthreads} multi-threading 
library.  The communication between the threads and services such IP sockets are 
provided by a custom C++ library wrapping the underlying C-based services.

A module may handle the readout for a specific subdetector or it
may analyze completed events for data quality monitoring or it may provide a
more peripheral service like logging messages to disk or providing a interface
between an external GUI and the DAQ core.  The two most important modules
in the DAQ are the {\it EventBuilder} and the {\it RunManager}, each of which
is described in more detail below.

Each module has a state variable which may be in one of four major
states: LOAD (just loaded), INIT (initializing, or initialized), RUN, or STOP
(stopping data acquisition or stopped).  Each of these states has substates
which indicate whether the process is ongoing (such as INIT/RUN), it has 
completed successfully (such as STOP/OK), or it has failed (such as INIT/FAIL).
Transitions between these states are triggered by broadcast messages. 

The most important method which modules use to communicate is broadcast 
messages.  Any module can
broadcast a very simple message (one integer number in the range 0-255) to 
all modules.  Messages in the range (0-31) can be ignored (masked off) if a 
module is not interested in them.  In practice, only a small fraction (16/256)
of the range of possible broadcast message codes was used.

A module can register one of its parameters as publicly accessible by assigning 
a string name and description to the parameter.  Other modules can query a 
given module for its list of defined parameters and their current values.  
They may also change parameters, if the exporting module has allowed this.
Within the DAQ, the set of parameters which were established when the INIT 
message was last broadcast are the set which are active for any data 
acquisition.

Modules which are directly involved in the high-bandwidth data-stream also 
exchange memory blocks containing event fragments or complete events as 
described below in the description of the EventBuilder.

\subsection{Configuration}

When the DAQ application starts up, it reads a configuration file which 
specifies the modules to load and gives initial parameter values for the modules.  
Only the EventBuilder, Reporter, and RunParamManager modules are loaded into 
the DAQ by default.  All other modules are loaded at run-time based on the
configuration file.  The configuration file is coded in XML and the DAQ uses
the libxml2\cite{libxml2} XML parser from the Gnome project.  
%An example XML configuration file is shown in Figure~{\ref{fig-XML}}.

The XML configuration file contains two sections.  The first section,
``$<$Modules$>$'', contains a list of modules to load.  For each module, the
name of the module within the DAQ is specified first, then the class name,
and the name of the shared library containing the class.  The DAQ code uses
a standard mechanism for loading classes and provides a preprocessor macro
%,``IMPLEMENT\_MODULE\_LOADER(CLASSNAME)'' 
to create the necessary stub code.
%\footnote{The IMPLEMENT\_MODULE\_LOADER(CLASSNAME) macro creates an {\sl extern ``C''} 
%function named \mbox{makeCLASSNAME()} which creates an instance of the module.}

The second section of file, ``$<$Parameters$>$'', contains a list of modules and provides
initial values for registered parameters in the module.  Any registered parameter
which is not read-only can be set using the parameter file.  The DAQ framework is
responsible for parsing the XML file and setting all the parameters specified in the
relevant modules before sending the LOAD broadcast command.

\subsection{The EventBuilder}

The EventBuilder module is responsible for collecting all the event fragments
from the subdetector readout modules and assembling complete valid events.
The EventBuilder defines streams of data associated with specific data blocks,
such as the normal HCALData block produced during beam data running or the 
WireChamberData block which contains the data from the wire chambers.  The
EventBuilder must be configured to know which blocks to expect in a given
run.  Only events with all the configured blocks are considered good.

The EventBuilder has a configurable property which is the number of
event buffers to allocate.  All readout modules read this parameter to decide
how many event fragment buffers they should locally allocate.  This is the upper bound
to how many events can be acquired within a spill.  The EventBuilder allocates
this many structures containing empty pointers for the readout streams.  As
readout modules receive triggers, they pass a pointer to the data for each
event to the EventBuilder which organizes the pointers by data stream and event
number.

Once the end of spill is signaled by the RunManager, the EventBuilder
starts trying to assemble complete events.  If all readout modules have 
completed (returned to the state RUN/RUN from RUN/BUSY), and there are still
events which do not have all required fragments, the spill is declared bad
and all events are thrown away.  This behavior is required since not all 
subdetectors have reliable event counters and this eliminates the possibility
of event data mismatch.  

If the spill is good, the events are distributed to the ``listeners''.  These
are threads (not full modules, but threads of a specific form set by the
EventBuilder) which have registered themselves with the EventBuilder to 
receive all good events.  The Writer module which stores the events to disk
uses such a listener thread.  Once all listeners have finished with the
data blocks, the EventBuilder sends a broadcast message to all readout
modules indicating that the event buffers are released and they can be
reused for the next spill.

\subsection{The RunManager}

The RunManager class is responsible for providing the TriggerData stream to
the EventBuilder, for generating Readout and End-Of-Spill broadcast messages,
and for logging completed runs into the database.  The base RunManager class
handles the database logging, but the other tasks are left to derived 
RunManager classes.  These derived classes generate trigger messages in 
different ways depending on the physical setup required.

The main beam-data manager is the H2Manager class, which interfaces to the H2
trigger using two CAENV513 NIM register VME boards.  The external trigger 
system signals the DAQ when triggered data should be read out and the
H2Manager signals back to the trigger when the DAQ is ready for the next event.
The trigger system also signals the manager when the spill is over and the
manager broadcasts the message through the DAQ.

Other RunManagers of note include the SourceManager which is used for taking 
radioactive source calibration data and the BenchManager which is used for 
LED pulser runs with the special pulser setup in H2 or for generating triggers
via software through the TTCvi module.

\subsection{HCAL Readout and Radioactive Source Data Taking}

The HCAL readout module is the most complicated of the readout modules.  
There are usually six or more HTRs in the readout crate, but not all HTRs are needed
at any given time.  The HCALReadout module allows different HTRs to be enabled or disabled
and handles the proper initialization for both the HTRs and the DCC.  

The HCALReadout module uses the SLINK driver developed by the University of 
Hawaii \cite{slinkhi}. The module configures the driver to put the data acquired during
a spill into a single DMA block in memory.  As the HTRs receive triggers and push the data
on to the DCC, the DCC forwards the data over the SLINK to the PC where it appears in the
DMA buffer.  At the end of a spill, the HCALReadout module picks through the buffer
and finds the event fragments from each HTR.  The DCC does not wrap the events with a 
header or trailer, so the DAQ simply takes ``one block from each HTR'' as a complete event.
If two blocks appear from one HTR before one from another, a warning message is issued.

During data taking with the radioactive source, no triggers are generated.  Instead, the
HTRs build histograms and forward them to the DCC at a fixed rate.  Within the DAQ, each
set of histograms is taken as ``one event''. Since the RunManager and EventBuilder need 
to know how many events were taken, the HCALReadout module counts how many it sees
in the DMA buffer when it receives a special broadcast message.  

The special message is sent by the SourceManager, which receives information about the 
position of the source and whether the run is complete via DIM (see below).  The 
SourceManager then asserts that the number of events which the HCALReadout saw 
occurred.  Since the reel position data arrives at a different rate than the 
histograms, the SourceManager then must map the reel data onto the same number of 
events as the histograms.  For this purpose the SourceManager uses the time-stamp which
is sent along with the reel data from the source control computer.

\subsection{Other Standard Modules}

Besides the major modules, there are many other modules which can be loaded into the 
DAQ.  
The ECALReadout module reads out the 7x7 ECAL matrix using a pair of CAENV792 VME QADCs.
The TDCReadout module reads out the wire chambers and other timing measurements
             using a CAENV767 VME TDC.  
%The relationships between readout modules and data streams are shown in Figure~\ref{fig-readout}.
The RunParamManager module provides central repository for information to be stored in the
                  RunData section of the files.  The module allows new parameters to be
                  created inside itself and all the parameters defined in the module
                  are automatically stored into the RunData header by the Writer.  Some
                  of the information is also put into the Runs Database. 
Finally, the Writer module is responsible for storing the completed events in a ROOT file.

\subsection{The Java GUI}

The DAQ process does not itself provide any mechanism for a user interface to control the run.  
Such an interface can provided by a module, and multiple modules could be written to control the 
DAQ using different communication techniques.  During normal operations, 
the DAQ was controlled using
a Java application which communicated with the DAQ through a custom remote-procedure-call (RPC) 
protocol.  The RPC protocol is transported using the standard TCP/IP transport, so the GUI program
does not have to be run on the same machine as the DAQ.  The Java GUI used the services of the ExternalInterface module.

The protocol allows the GUI to obtain the state of the system in the form of an XML 
document which has the same format and structure as the ``$<$Parameters$>$'' section of
the configuration file.  The GUI parses this XML document to determine the state of all the
registered parameters.  The protocol also allows the GUI to change the values of parameters and
to send broadcast command messages.  
%A screen-shot of the DAQ GUI in action is shown in Figure~\ref{fig-screenshot}. 
The Java GUI also incorporates a plotting package which can access the histograms exported by 
the StatsMan module through the HistoExporter class.  These histograms are available on an IP port in
text form, and the GUI can plot these online monitoring histograms.  
%Some examples of plots from the online monitoring are shown in Figure~\ref{fig-onlinemonitor}.

\subsection{Interaction with DIM}

The Distributed Information Management System software \cite{dim} has emerged as a
powerful, general, and freely-distributable tool for exporting and accessing 
structured data and sending commands.  Heavy use of DIM will be made at CMS and LHC, not
least because of the DIM$\leftrightarrow$PVSS gateway which has been developed.  The
current version of the DAQ does not include a module to allow direct control of the DAQ
through DIM, but it does use DIM to gather several important types of data.

The DAQ includes a module called InfoGrabber which allows the DAQ to pull information from
several DIM servers.  
From the BeamInfoDataServer, the DAQ can learn the beam energy, beam settings, and 
the particle type.  From the TableDataServer, the DAQ can learn the position of the 
movable table on which the HCAL and ECAL are mounted.  Other servers provide temperature
measurements of the ECAL matrix, and allow configuration commands to be sent to the
QIEs and other front-end chips from a Windows machine.
DIM is also used during source data taking to obtain the reel position and other
data from the source control computer.
 
\subsection{Persistent Data Format}

The data is stored in a ROOT Tree format, where each event is represented by an object of type
TestBeamData, which may contain pointers to other data blocks.  Each of the data blocks is placed
onto its own ROOT branch.  A C++ library is provided for users to access and reconstruct the data
in the ROOT files.  Complete documentation for this library can be found in reference \cite{HTBDAQData}.

\subsection{Runs Database}

Basic configuration data and statistics for each saved run are stored in a runs database.  The database
consists of a set of XML files which contain entries for a range of run numbers.  This database is
cached in AFS space and also is available for querying on the web.  More information about the
runs database can be found at \cite{HTBDAQRunsdb} and the main DAQ website \cite{tbpage}.

\section{Performance of the DAQ Software}

The DAQ Software worked well for the 2002 testbeam.  During the summer running, 3,081 runs were taken
covering some 53,182 spills and containing 100,655,201 events.  The data volume recorded into the
central data recording was 280GB, with an average compression ratio (from ROOT) of 2.1:1.  On average, the
DAQ was run at rate of 1700 events per spill or 350 Hz, but when the beam intensity was increased the readout
rate reached 4720 events per spill or 960 Hz.  The data acquired has proved
extremely useful in understanding the performance of the final system and
testing the calibration and resolution of the calorimeter.

After the summer testbeam, the DAQ software was used at Fermilab to evaluate new QIE boards for the 2003 test beam
and to acquire data on radiation damage of QIEs at the Indiana Cyclotron Facility.

The testbeam program for the HCAL will continue in 2003, and the DAQ software
is changing to approach the final software structure which will be used in
CMS.  The custom base library will be replaced by the xDAQ library \cite{xdaq},
while the configuration management will be significantly updated to support
increasing complicated configurations.  The custom XML-based runs database will
be transferred into a standard SQL server form, which will improve the 
stability of the system.  The goals for 2003 include structured beam running
for the barrel part of the calorimeter, a first look at the endcap sections 
with the final electronics, and study and calibration of the forward hadron
calorimeter.  It will be challenging to acquire and process the required data
volumes, but the results will be bring the CMS HCAL closer to operational 
status.


\begin{thebibliography}{9}   % Use for  1-9  references
%\begin{thebibliography}{99} % Use for 10-99 references

%\bibitem{accelconf-ref}
%http://www.cern.ch/accelconf

 \bibitem {htr} T. Grassi. ``CMS HCAL Trigger/Readout working page.'' {\bf http://tgrassi.home.cern.ch/tgrassi/hcal}.
 \bibitem {dcc}	E. Hazen. ``CMS HCAL Data Concentrator working page.'' {\bf http://ohm.bu.edu/~hazen/hcal}.
 \bibitem {pthreads} B.~Nichols, D.~Buttlar, and J.~Farrell. {\bf Pthreads programming}, Nutshell handbook.
 \bibitem {libxml2} Daniel Veillard. ``The XML C library for Gnome.'' {\bf http://xmlsoft.org}.
 \bibitem {slinkhi} S.~Isani, Canada-France-Hawaii Telescope group. {\bf http://software.cfht.hawaii.edu/sspci}.
 \bibitem {dim} DIM, a Portable, Light Weight Package for Information Publishing, Data Transfer and Inter-process Communication (pdf) Presented at:
     International Conference on Computing in High Energy and Nuclear Physics (Padova, Italy, 1-11 February 2000), \\
                DIM working site. {\bf http://dim.home.cern.ch/dim}.
 \bibitem {tbpage} HCAL 2002 Testbeam working page. {\bf http://cern.ch/cms-testbeamh2}.
 \bibitem {HTBDAQ} J. Mans. ``HTBDAQ.'' {\bf http://flywheel.princeton.edu/~jmmans/HTBDAQ}.
 \bibitem {HTBDAQData} J. Mans. ``HTBDAQ\_Data Library.'' {\bf http://flywheel.princeton.edu/~jmmans/HTBDAQ\_Data}.
 \bibitem {HTBDAQRunsdb} J. Mans. ``HTBDAQ\_RunsDB Library and Utilities.'' {\bf http://flywheel.princeton.edu/~jmmans/HTBDAQ\_rundb}.
 \bibitem {xdaq} xDAQ working page. {\bf http://cern.ch/xdaq/ }.
%\bibitem{templates-ref}
%http://www.cern.ch/accelconf/templates.html

\end{thebibliography}

\end{document}